\documentclass[prl,aps,twocolumn,showpacs,superscriptaddress]{revtex4}

\usepackage{graphicx}
\usepackage{dcolumn}
\usepackage{bm}

\begin{document}


\title{Search for Exotic Baryons in 800 GeV $pp\rightarrow p\Xi^\pm\pi^\pm X$}

\author{D.C. Christian}
\affiliation{Fermi National Accelerator Laboratory, Batavia, IL}

\author{J. Felix}
\affiliation{Universidad de Guanajuato, Le\'on, Guanajuato, M\'exico}

\author{E.E. Gottschalk}
\affiliation{Fermi National Accelerator Laboratory, Batavia, IL}
\author{G. Gutierrez}
\affiliation{Fermi National Accelerator Laboratory, Batavia, IL}

\author{E.P. Hartouni}
\affiliation{Lawrence Livermore National Laboratory, Livermore, CA}

\author{B.C. Knapp}
\affiliation{Columbia University, Nevis Laboratory, Irvington, NY}

\author{M.N. Kreisler}
\affiliation{Lawrence Livermore National Laboratory, Livermore, CA}
\affiliation{University of Massachusetts, Amherst, MA}

\author{G. Moreno}
\affiliation{Universidad de Guanajuato, Le\'on, Guanajuato, M\'exico}

\author{M.A. Reyes}
\affiliation{Universidad de Guanajuato, Le\'on, Guanajuato, M\'exico}
\author{M. Sosa}
\affiliation{Universidad de Guanajuato, Le\'on, Guanajuato, M\'exico}

\author{M.H.L.S. Wang}
\affiliation{Fermi National Accelerator Laboratory, Batavia, IL}
\author{A. Wehmann}
\affiliation{Fermi National Accelerator Laboratory, Batavia, IL}

\date{\today}

\begin{abstract}
We report the results of a high-statistics, sensitive search for narrow baryon resonances
decaying to $\Xi^-\pi^-$, $\Xi^-\pi^+$, $\overline{\Xi}^+\pi^-$, and $\overline{\Xi}^+\pi^+$.
The only resonances observed are the well known $\Xi^0(1530)$ and $\overline{\Xi^0}(1530)$.
No evidence is found for a state near 1862 MeV, previously reported by NA49\cite{NA49}.
At the $95\%$ confidence level, we find
the upper limit for the production of a Gaussian enhancement with $\sigma=7.6$ MeV
in the $\Xi^-\pi^-$ effective mass spectrum to be $0.3\%$ of the number of observed
$\Xi^0(1530)\rightarrow\Xi^-\pi^+$.  We find similarly restrictive upper limits for an 
enhancement at 1862 MeV in the $\Xi^-\pi^+$, $\overline{\Xi}^+\pi^-$, and $\overline{\Xi}^+\pi^+$ mass spectra.
\end{abstract}

\maketitle


A number of different experiments have reported evidence for
the existence of the $\Theta^+$, a strangeness +1 baryon that decays to $nK^+$ or $pK^0$
\cite{leps}\cite{dian}\cite{clas}\cite{safr}\cite{HT}\cite{ZT}\cite{COSY}\cite{CT}.
The $\Theta^+$ mass is reported to be
approximately 1540 MeV and its width less than $\sim$20 MeV.
The $\Theta^+$ has been interpreted as a pentaquark, consisting of two up quarks,
two down quarks, and one anti-strange quark.
If the pentaquark interpretation is correct, then a large number of similar states are
expected\cite{multiquark}.
Many of these states besides the $\Theta^+$ have quantum numbers not possible
for baryons composed only of three quarks.  One such state is a charge -2, strangeness +2 baryon expected
to decay to $\Xi^-\pi^-$.  In 2003, Wilczek and Jaffe predicted that this state should
have a mass of approximately 1750 MeV and should be narrow, with a width 
only $\sim50\%$ greater than the width of the $\Theta^+$\cite{Jaffe}.
Shortly after this prediction, NA49 reported evidence for a $\Xi^-\pi^-$ resonance 
produced in proton-proton interactions at $\sqrt{s}=17.2$ GeV with a mass of
1862 MeV and width less than the detector resolution of 18 MeV FWHM\cite{NA49}.
NA49 also reported evidence for states with similar masses and widths decaying to $\Xi^-\pi^+$,
$\overline{\Xi}^+\pi^-$, and $\overline{\Xi}^+\pi^+$.

However, the experimental case for the existence of pentaquarks is not yet compelling.
Although the $\Theta^+$ has been observed in a variety of different reactions,
many experiments have failed to confirm its existence
\cite{HERA-B}\cite{Hyp}\cite{ST}\cite{BES}\cite{ALEPH}. 
Futhermore, the HERA-B collaboration has reported that they do not observe an enhancement at 1862 MeV
in $\Xi^-\pi^-$ produced near Feynman $x=0$ in interactions of 920 GeV protons with a variety
of nuclear targets\cite{HERA-B}. A search in $\Sigma^-$-nucleus collisions at 350 GeV by
WA89\cite{WA89} yielded only an upper limit on the production of a $\Xi^-\pi^-$ resonance.
Similarly, the HERMES collaboration\cite{HERMES} searching in photoproduction,
the ZEUS\cite{ZEUS} collaboration searching in deep inelastic $ep$ collisions,
and the COMPASS collaboration\cite{COMPASS} searching in deep inelastic $\mu^+N$ collisions
were also only able to place limits on the production of this state.

\begin{figure}
\label{fig:jggspec}
\includegraphics{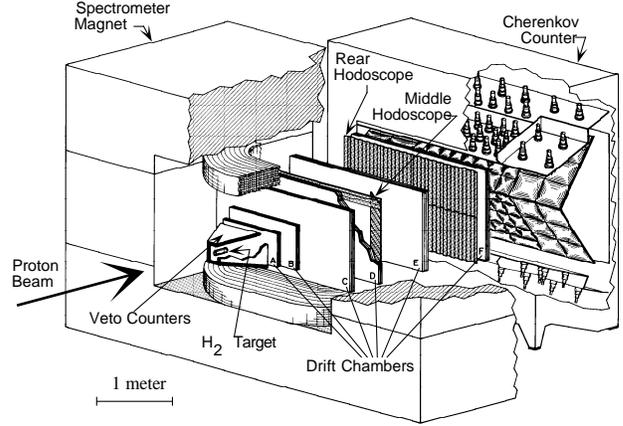}
\caption{\it E690 Multiparticle Spectrometer.}
\end{figure}

We have conducted a high-statistics, sensitive search for new resonances decaying to
$\Xi^-\pi^-$, $\Xi^-\pi^+$, $\overline{\Xi}^+\pi^-$, and $\overline{\Xi}^+\pi^+$ using data collected in 1991 by
Fermilab E690.
The E690 apparatus consisted of an open-geometry multiparticle
spectrometer (shown in Fig. 1)
and a beam spectrometer.  The multiparticle
spectrometer was used to measure the target system ($X$) in reactions
\begin{equation}
pp \to p_{fast}X.
\end{equation}
The beam spectrometer was used to measure the incident 800 GeV beam and
scattered proton.  A liquid hydrogen target was located just upstream of the
multiparticle spectrometer.
The multiparticle spectrometer included six
mini-drift proportional wire chambers, each with four signal planes.
Five of the wire chambers were located inside a large aperture dipole magnet,
and the sixth chamber was located just downstream of the magnet.
The magnet had a central field of $\sim$7 kG.  The angular coverage was $\pm 580$ mrad
in the horizontal (bend) direction and $\pm 410$ mrad in the vertical direction.
The spaces between the wire chambers were filled with helium to reduce multiple scattering.
Particle identification was provided by 102 time-of-flight scintillation counters and a 96-cell
Freon-114 Cherenkov counter with a pion momentum threshold of 2.57 GeV.

Approximately 4.3 billion events were recorded using a trigger that imposed two event
selection requirements.  The trigger required:

\begin{enumerate}
\item{An interaction in the target region, indicated by signals from one or more time-of-flight
scintillation counters.}
\item{A scattered beam particle, detected by counters outside of the beam spot in the forward
beam spectrometer.}
\end{enumerate}

Event reconstruction was performed for the entire data sample.  The primary (interaction)
vertex was constrained to lie on an incoming beam-track trajectory.
Events without an identified primary vertex were not processed further.
For events in which evidence of a ``vee'' or ``cascade'' decay was found,
the tracks were refit with the constraint that each daughter vertex ``point back''
to its parent, and the daughter was then ``assigned'' to the parent.  Mass constrained fits were not performed. 
The following decay types were searched for:  $\gamma \rightarrow e^+e^-$,
$K^0_s \rightarrow \pi^+\pi^-$, $\Lambda \rightarrow p\pi^-$,
$\overline{\Lambda} \rightarrow \overline{p}\pi^+$, $\Xi^- \rightarrow \Lambda\pi^-$,
$\overline{\Xi}^+ \rightarrow \overline{\Lambda}\pi^+$,
$\Omega^- \rightarrow \Lambda K^-$, $\overline{\Omega}^+ \rightarrow \overline{\Lambda} K^+$, 
$K^+ \rightarrow \pi^+\pi^+\pi^-$, and $K^- \rightarrow \pi^-\pi^-\pi^+$.
Any remaining vertices were assumed to be caused by secondary interactions or interactions
of spectator beam particles.

For this analysis, events were selected if they had a candidate $\Xi^-$ or $\overline{\Xi}^+$ decay.
The sample contains just over 500,000 unambiguously identified $\Xi^-$'s and approximately 34,500 $\Xi^-$
candidates that may be interpreted as $\Omega^-$ if the negative track is assumed to be a
$K^-$ rather than a $\pi^-$.  Of these, approximately 513,000 are assigned to the primary
vertex.  The sample contains just over 160,000 unambiguously identified $\overline{\Xi}^+$ candidates and
approximately 9,600 ambiguous $\overline{\Xi}^+$/$\overline{\Omega}^+$ candidates.  Of these, approximately
153,700 are assigned to a primary vertex.  The effective mass spectra of the $\Xi^-$
and $\overline{\Xi}^+$ candidates assigned to a primary vertex are shown in Fig. 2.
There is slightly more background for the $\overline{\Xi}^+$ distribution compared to the $\Xi^-$
distribution.  The effective mass spectra of the $\Lambda$ and $\overline{\Lambda}$ candidates
contained in the $\Xi$ candidates are also shown in Fig. 2.

\begin{figure}[hbt]
\label{fig:xilammass}
\includegraphics[width=7.5cm]{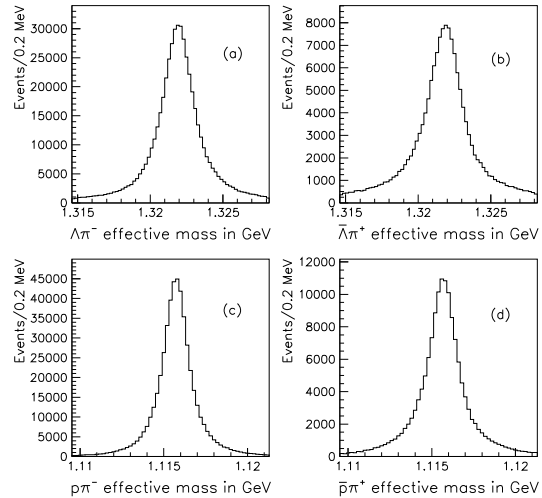}
\caption{\it Hyperon effective mass spectra in events used for this analysis:
a)$\Xi^-$, b)$\overline{\Xi}^+$, c)$\Lambda$ from $\Xi^-$ decay, and d)$\overline{\Lambda}$ from $\overline{\Xi}^+$ decay.
The full width at half maximum is $1.8$ MeV for the $\Lambda$ and $\overline{\Lambda}$
and $2.6$ MeV for the $\Xi^-$ and $\overline{\Xi}^+$.}
\end{figure}

\begin{figure}[hbt]
\label{fig:xipimass}
\includegraphics[width=7.5cm]{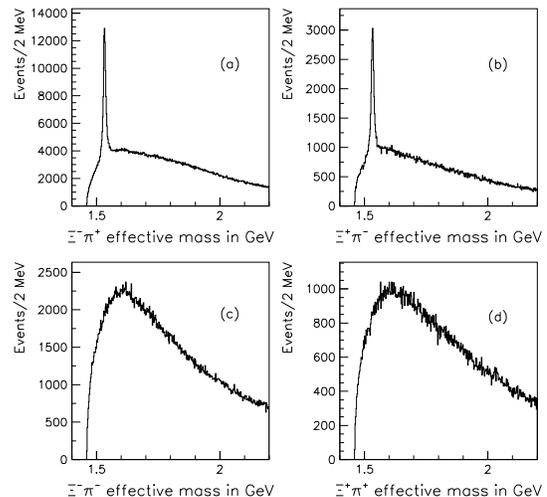}
\caption{\it Effective mass spectra for: a)$\Xi^-\pi^+$, b)$\overline{\Xi}^+\pi^-$,
c)$\Xi^-\pi^-$, and d)$\overline{\Xi}^+\pi^+$.}
\end{figure}

\begin{figure*}[hbt]
\label{fig:xi1530mass}
\includegraphics[width=15cm]{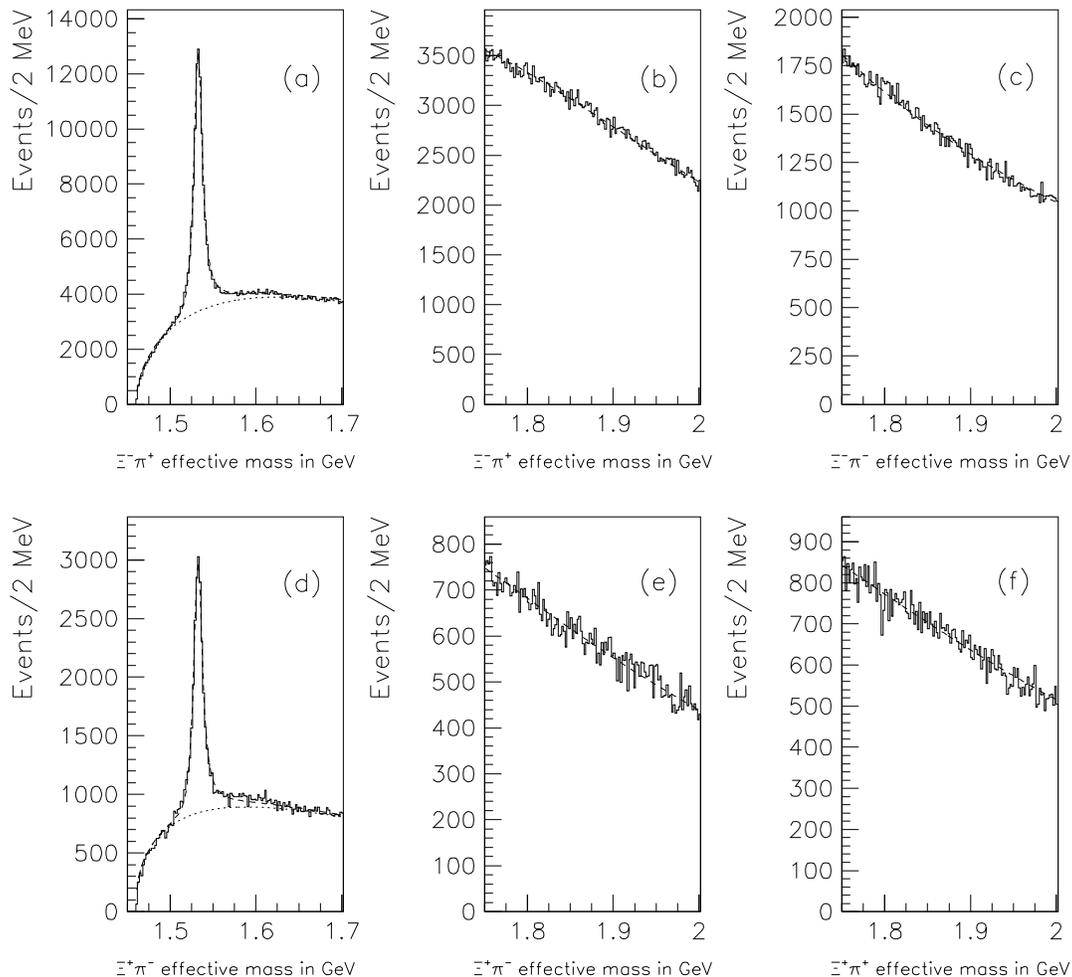}
\caption{\it Fits to the $\Xi\pi$ effective mass spectra:  a) and d) show the region near 1530 MeV.
The superimposed curves show the background function and the sum of the background and a relativistic
Breit-Wigner resonance, as described in the text.  b), c), e), and f) show the region near 1862 MeV.
The superimposed curves consist of the background function only.}
\end{figure*}

$\Xi\pi$ mass distributions were computed by pairing each $\Xi$ candidate that was
assigned to a primary vertex with every charged track (assumed
to be a pion) assigned to the same primary vertex.
Direct particle identification was not used in this analysis because a significant fraction
of the tracks had momentum too high to provide good time-of-flight particle identification
and too low to allow Cherenkov identification.  The effective mass distributions
for the four charge combinations are shown in Fig. 3.  The $\Xi^-\pi^+$ and $\overline{\Xi}^+\pi^-$
mass distributions contain prominent $\Xi^0(1530)$ 
and $\overline{\Xi}^0(1530)$ resonances.
None of the four effective mass distributions contains any other obvious resonance.

The plots in Fig. 3 clearly show that if a narrow pentaquark decaying to $\Xi\pi$ 
exists, then its production is highly suppressed with respect to the
production of $\Xi^0(1530)$ and $\overline{\Xi}^0(1530)$ in these data.
In particular, the data contain no evidence for the existence of a pentaquark with mass 1862 MeV
in any of the four final states being studied.
In order to quantify this conclusion,
we computed $95\%$ confidence level upper limits\cite{FeldmanCousins}
for Gaussian signals with mass = 1862 MeV and $\sigma$=7.6 MeV (the NA49 mass resolution).

For comparison, we estimated the yield of $\Xi^0(1530)$ and $\overline{\Xi}^0(1530)$.
We fit the $\Xi^-\pi^+$ and $\overline{\Xi}^+\pi^-$ effective mass distributions between threshold and 2200 MeV.
The mass distributions were fit to the sum of a background function
and a signal consisting of a relativistic ($L=1$) Breit-Wigner function\cite{Jackson}
convoluted with a Gaussian with $\sigma$ = 2.5 MeV.
The Gaussian smearing term represents the spectrometer mass resolution for effective mass close
to 1530 MeV.  It was set so that the fit value of the
widths of the $\Xi^0(1530)$ and $\overline{\Xi}^0(1530)$ approximately agreed
with the values listed by the Particle Data Group.  The inferred mass resolution of 2.5 MeV is
consistent with the observed width of other particles, such as $\Xi^-$ and $\Lambda^0$.
The form of the background function was 
$F(M)=(M-M_T)^{\alpha}e^{\beta(M-M_T)}$, where $M_T$ is the $\Xi\pi$ threshold mass of
1.46 GeV$^2$ and $\alpha$ and $\beta$ are fit parameters.
This background function was also used in reference\cite{LASS}.

The results of the fits to the $\Xi^-\pi^+$ and $\overline{\Xi}^+\pi^-$ effective mass distributions
are shown in Fig. 4 and summarized in Table~\ref{table:cascade1530}
\footnote{The Particle Data Group value (given in S. Eidelman {\it et al.},
Physics Letters {\bf B592}, 1 (2004)) for the
mass of the $\Xi(1530)$ is $1531.8 \pm 0.3$ MeV.
Our fit masses for the $\Xi^0(1530)$ and $\overline{\Xi}^0(1530)$ are approximately
1 MeV higher than the PDG value.  However,
we have not corrected our measurements for drifts of the magnetic field during data
taking and have not evaluated the systematic error in our mass measurements.}.
The doubly charged $\Xi^-\pi^-$ and $\overline{\Xi}^+\pi^+$ effective mass distributions
were fit satisfactorily using only the background function described above.
The upper limits computed for a Gaussian enhancement at 1862 MeV
with $\sigma = 7.6$ MeV are given in Table~\ref{table:exotic1862}.

\begin{table}[hbt]
\caption{\it $\Xi^0(1530)$ and $\overline{\Xi}^0(1530)$ fit results.
The errors shown are statistical errors only.}
\label{table:cascade1530}
\begin{tabular}{|l|l|l|l|}
\hline
Decay Mode & $\Xi^-\pi^+$ & $\overline{\Xi}^+\pi^-$\\
\hline
$\chi^2/dof$& 683/364 & 443/364\\
Fit Mass (MeV)& 1532.73 $\pm$ 0.03 & 1532.65 $\pm$ 0.06 \\
Fit Width (MeV)&8.96 $\pm$ 0.06 &9.41 $\pm$ 0.01\\
Yield& 93728 $\pm$ 422& 22211 $\pm$ 219\\ 
\hline
\end{tabular}\\[2pt]
\end{table}

\begin{table}[hbt]
\caption{\it $95\%$ confidence level upper limits for a signal at 1862 MeV. Each limit is also
expressed as a fraction of the $\Xi^0(1530)$ yield (for baryon modes) or the
$\overline{\Xi}^0(1530)$ yield (for antibaryon modes).}
\label{table:exotic1862}
\begin{tabular}{|c|c|c|}
\hline
Decay Mode & Limit on Gaussian signal (events)\\
           & (Mass = 1862 MeV, $\sigma = 7.6$ MeV)\\
\hline
$\Xi^-\pi^+$ & 1020 $(1.1\%)$\\
$\Xi^-\pi^-$ & 310 $(0.3\%)$\\
$\overline{\Xi}^+\pi^-$ & 290 $(1.3\%)$\\
$\overline{\Xi}^+\pi^+$ & 288 $(1.3\%)$\\
\hline
\end{tabular}\\[2pt]
\end{table}

In summary, we have performed a study of inclusive $\Xi^\pm\pi^\pm$ production in beam diffraction
reactions of the type $pp \rightarrow p_{fast}X$ with an 800 GeV proton beam and a liquid hydrogen
target.  Strong signals are observed for both the $\Xi^0(1530)$ and the $\overline{\Xi}^0(1530)$.  No
other peak is observed in any of the four $\Xi\pi$ effective mass distributions.
If a Gaussian line shape with $\sigma=7.6$ MeV is assumed, the number of
$\Xi^-\pi^-$ produced at 1862 MeV is less than $0.3\%$ of the observed number of
$\Xi^0(1530)\rightarrow\Xi^-\pi^+$.  The limit for the $\Xi^-\pi^+$ final state is $1.1\%$
of the $\Xi^0(1530)$ yield.
The limits for the $\overline{\Xi}^+\pi^-$ and $\overline{\Xi}^+\pi^+$ final states are $1.3\%$ of the observed number
of $\overline{\Xi}^0(1530)\rightarrow\overline{\Xi}^+\pi^-$.  No evidence is found for a state near 1862 MeV.

We acknowledge the superb efforts by the staffs
at the University of Massachusetts, Columbia University,
Fermilab, and Lawrence Livermore National Laboratory. 
This work was supported in part by National Science Foundation Grants
No. PHY90-14879 and No. PHY89-21320, by the Department of Energy
Contracts No. DE-AC02-76 CHO3000, No. DE-AS05-87ER40356 and
No. W-7405-ENG-48, and by CoNaCyT of M\'exico under Grants 2002-C01-39941, NSF-J200.810, and
CONCYTEG-03-16-K118-024.

\end{document}